\documentclass[11pt,twoside]{article}

\usepackage{asp2014}

\aspSuppressVolSlug
\resetcounters

\begin{document}

\title{Optical Variability of "Light-weight" Supermassive Black Holes at a Few Percent Level from ZTF Forced-Photometry Light Curves}

\author{Mariia~Demianenko,$^{1,2,3}$ Kirill~Grishin,$^2$ Victoria~Toptun,$^{2,5}$ Igor~Chilingarian,$^{4,2}$ Ivan~Katkov,$^{6,7,2}$ Vladimir~Goradzhanov,$^{2,5}$ and Ivan~Kuzmin$^{2,5}$}
\affil{$^1$Moscow Institute of Physics and Technology (National Research University), Moscow, Russia; \email{demyanenko.mv@phystech.edu}}
\affil{$^2$Sternberg Astronomical Institute, M.V. Lomonosov Moscow State University, Moscow, Russia}
\affil{$^3$HSE University, Moscow, Russia}
\affil{$^4$Center for Astrophysics -- Harvard and Smithsonian, Cambridge, USA}
\affil{$^5$Department of Physics, M.V. Lomonosov Moscow State University}
\affil{$^6$New York University Abu Dhabi, UAE}
\affil{$^7$Center for Astro, Particle, and Planetary Physics, NYU Abu Dhabi, UAE}

\paperauthor{Mariia~Demianenko}{demyanenko.mv@phystech.edu}{0000-0002-8297-6386}{Author1 Institution}{Author1 Department}{City}{State/Province}{Postal Code}{Country}
\paperauthor{Kirill~Grishin}{grishin@voxastro.org}{0000-0003-3255-7340}{Author2 Institution}{Author2 Department}{City}{State/Province}{Postal Code}{Country}
\paperauthor{Victoria~Toptun}{victoria.toptun@voxastro.org}{0000-0003-3599-3877}{Author3 Institution}{Author3 Department}{City}{State/Province}{Postal Code}{Country}
\paperauthor{Igor~Chilingarian}{igor.chilingarian@cfa.harvard.edu}{ORCID_Or_Blank}{Author3 Institution}{Author3 Department}{City}{State/Province}{Postal Code}{Country}
\paperauthor{Ivan~Katkov}{katkov.ivan@gmail.com}{0000-0002-6425-6879}{Author3 Institution}{Author3 Department}{City}{State/Province}{Postal Code}{Country}
\paperauthor{Vladimir~Goradzhanov}{goradzhanov.vs17@physics.msu.ru}{0000-0002-2550-2520}{Author3 Institution}{Author3 Department}{City}{State/Province}{Postal Code}{Country}
\paperauthor{Ivan~Kuzmin}{kuzmin.ia19@physics.msu.ru}{0000-0003-2888-2474}{Author3 Institution}{Author3 Department}{City}{State/Province}{Postal Code}{Country}



\begin{abstract}
Large time-domain surveys provide a unique opportunity to detect and explore variability of millions of sources on timescales from days to years. Broadband photometric variability can be used as the key selection criteria for weak type-I active galactic nuclei (AGN), when other ``direct'' confirmation criteria like X-ray or radio emission are unavailable. However, to detect variability of rather weak AGN powered by intermediate-mass black holes, typical sensitivity provided by existing light curve databases is insufficient. Here we present an algorithm for post-processing of light curves for sources with stochastic variability, retrieved from the The Zwicky Transient Facility (ZTF) Forced Photometry service. Using our approach, we can filter out spurious data points related to data reduction artefacts and also eliminate long-term trends related to imperfect photometric calibration. We can now confidently detect the broad-band variability at the 1-3~\%\ level which can potentially be used as a substitute for expensive X-ray follow-up observations.
\end{abstract}

\section{Introduction and Motivation}
An important property of active galactic nuclei (AGN) is flux variability on all timescales at all wavelengths. The shortest timescales are defined solely by the black hole mass $M_{BH}$ and no variability is expected at timescales $\lesssim GM_{BH}/c^3$; intermediate timescales (hours to weeks) are presumably driven by as yet unidentified processes within the disk and corona, and bracketed by the dust torus size around the AGN, which is proportional to X-ray luminosity $L_X^{0.5}$, which is also related to $M_{BH}$ via the Eddington limit; the longer timescales (years) reflect the matter distribution in the nuclear region of a galaxy, e.g. the size and mass of the nuclear star cluster; finally, the AGN duty cycle depends on the availability of the gas supply that feeds the BH and can become totally unrelated to the BH itself. Because of the mass dependence of shorter timescales, intermediate-mass black holes (IMBHs; $M_{BH}<10^5 M_{\odot}$) work as a ``time-squeezing machine'': the shortest timescales change $\propto M_{BH}$, so 1 year for a $7\times10^4 M_{\odot}$ IMBH corresponds to a few centuries for a typical (2--5)$\times10^7 M_{\odot}$ super-massive black hole (SMBH). At the same time, the variability properties of IMBHs and ``light-weight'' SMBHs ($M_{BH}<10^6 M_{\odot}$) remain completely unexplored. 

Optical variability analysis can select and confirm many more ``light-weight'' SMBH and IMBH candidates, because X-ray follow-up observations cannot be performed for large samples of objects with high enough time cadence. ZTF Data Release \citep{2019PASP..131a8003M} aperture light curves cannot be used for unambiguous AGN confirmation because the systematic uncertainties reaching 15-20$\%$ make the detection of variability of weak AGN powered by ``light-weight'' SMBHs unfeasible. Instead, we build more reliable light curves from difference images using the ZTF Forced Photometry service to detect optical variability at a few percent level from a large sample weak AGN selected from archival X-ray observations and H$\alpha$ broad line width. Still, to tackle the systematic errors related to zero-point calibration issues, we need additional post-processing if the desired level of uncertainty is a few percent.

\citet{2018ApJ...863....1C} identified a sample of IMBHs ($3 \times 10^4 < M_{BH} < 2 \times 10^5 M_{\odot}$) by applying the {\sc NBursts} full spectral fitting \citep{2007IAUS..241..175C,2007MNRAS.376.1033C} to SDSS and SDSS/eBOSS optical spectra in the RCSED catalog \citep{2017ApJS..228...14C} with the virial $M_{BH}$ calibration \citep{2004ApJ...610..722G,optic_fm}. For our analysis we use a part of this list complemented with new sample of ``light-weight'' SMBHs selected in a similar fashion, for which we found X-Ray counterparts in archival data from XMM-Newton, Chandra, and Swift observatories. To illustrate how our algorithm works, we use 2 objects: (i) J163159.59+243740.2 is a candidate binary ``light-weight'' SMBH with a distinct variability seen even in the ZTF Data Release light curves; (ii) J152442.58+292701.7 is an X-ray confirmed ``light-weight'' SMBH with clearly detected broad H$\alpha$ lines in its optical spectra.

\section{Methods}
\subsection{Filtering spurious outliers}
We use the guidelines for the Forced Photometry service to filter by the following parameters: signal-to-noise ratio, distance to the nearest reference source, pixel quality indicators to remove observations with uncorrected cosmic ray hits and hot pixels in scientific images, sharpness values, and the {\sc nearestrefchi} parameter of the reference source to avoid spikes or hard edges in extended reference sources \citep{2019PASP..131a8003M}.
According to the available information, a large fraction of ZTF data was acquired in non-photometric conditions (e.g. scattered clouds, variable transparency). Therefore, outliers in light curves might be artefacts originating from data analysis rather than real values reflecting physical variability. Hence, a formally computed statistical criterion of variability becomes unreliable unless spurious measurements have been removed.

\articlefigurefour{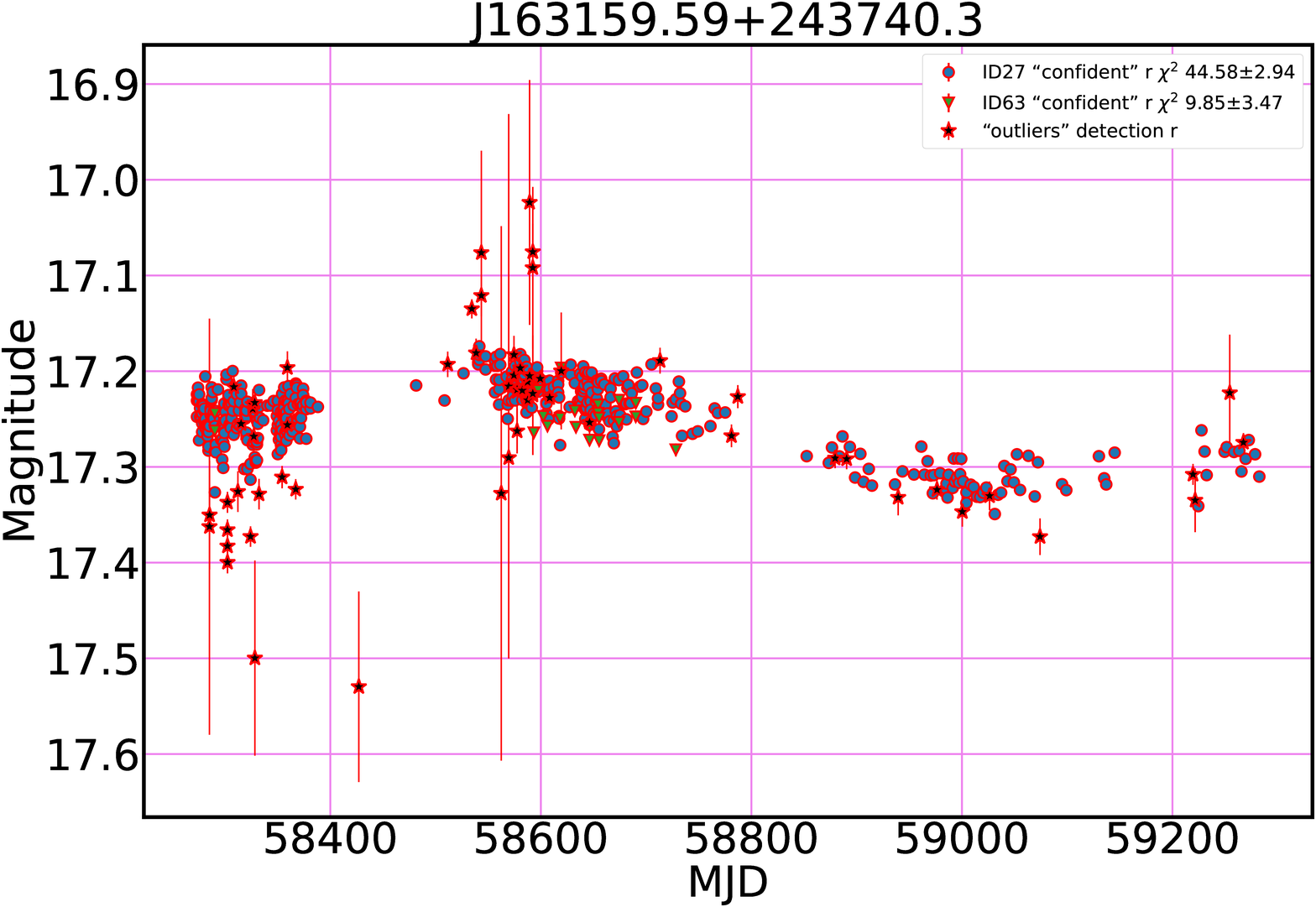}{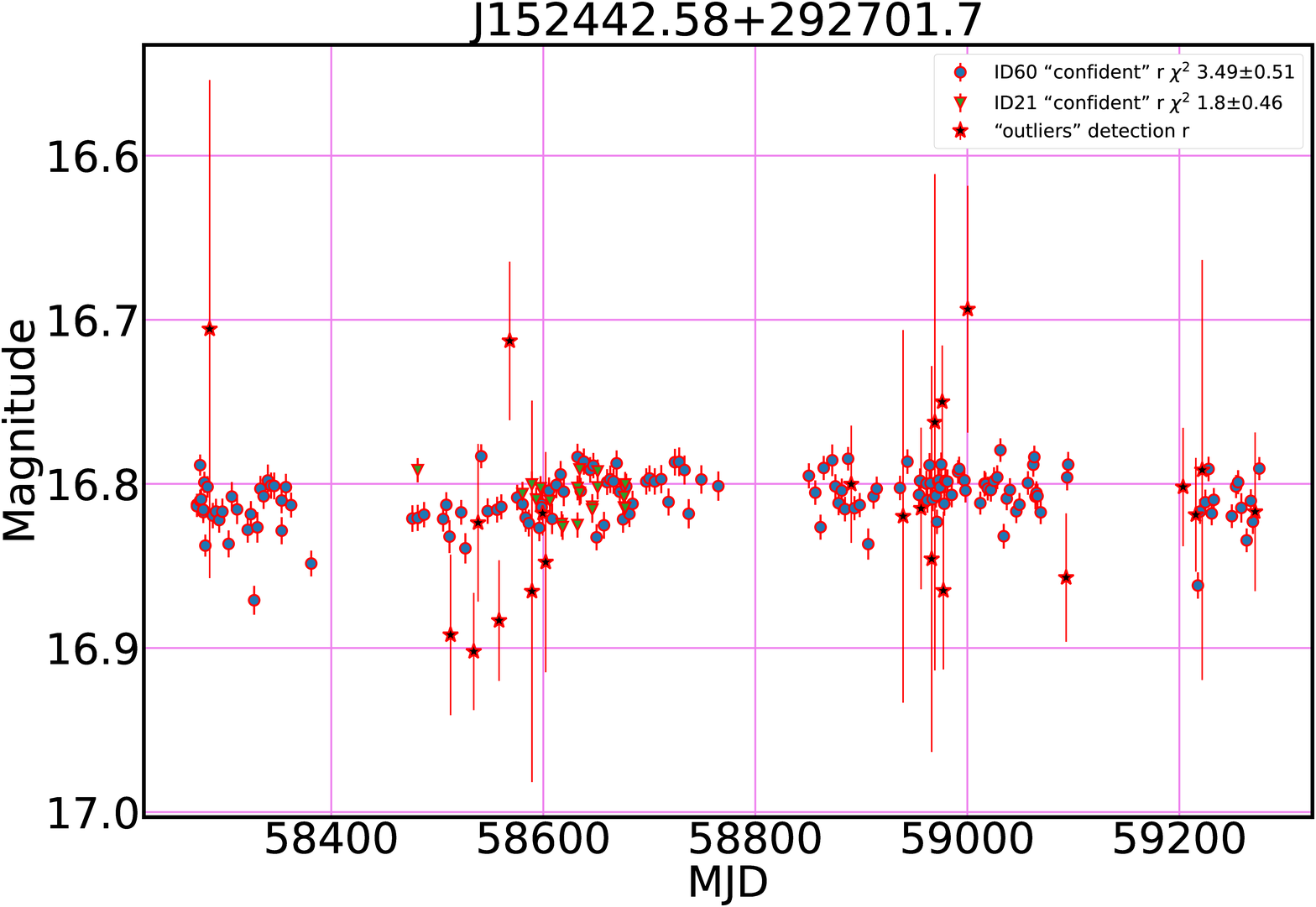}{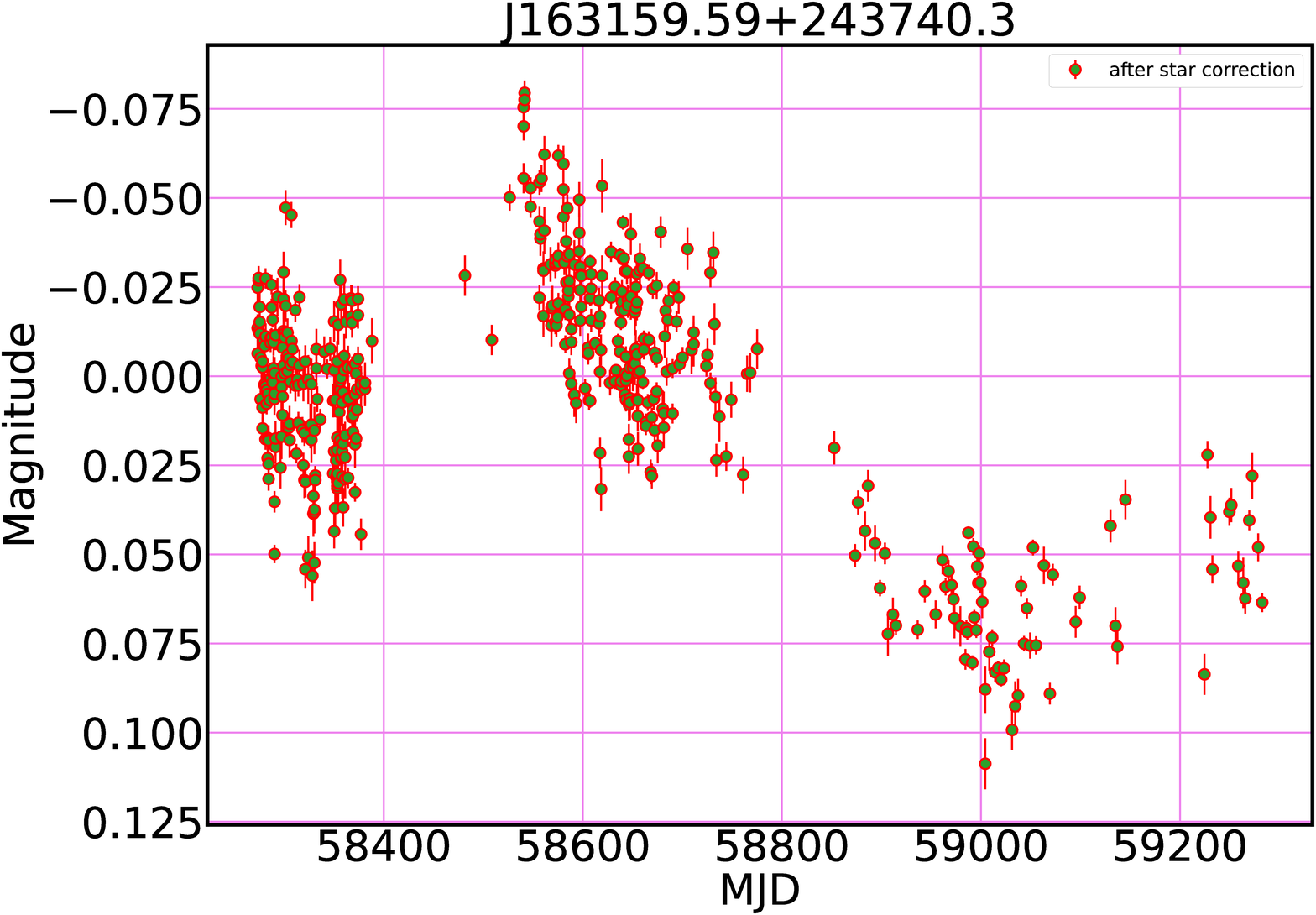}{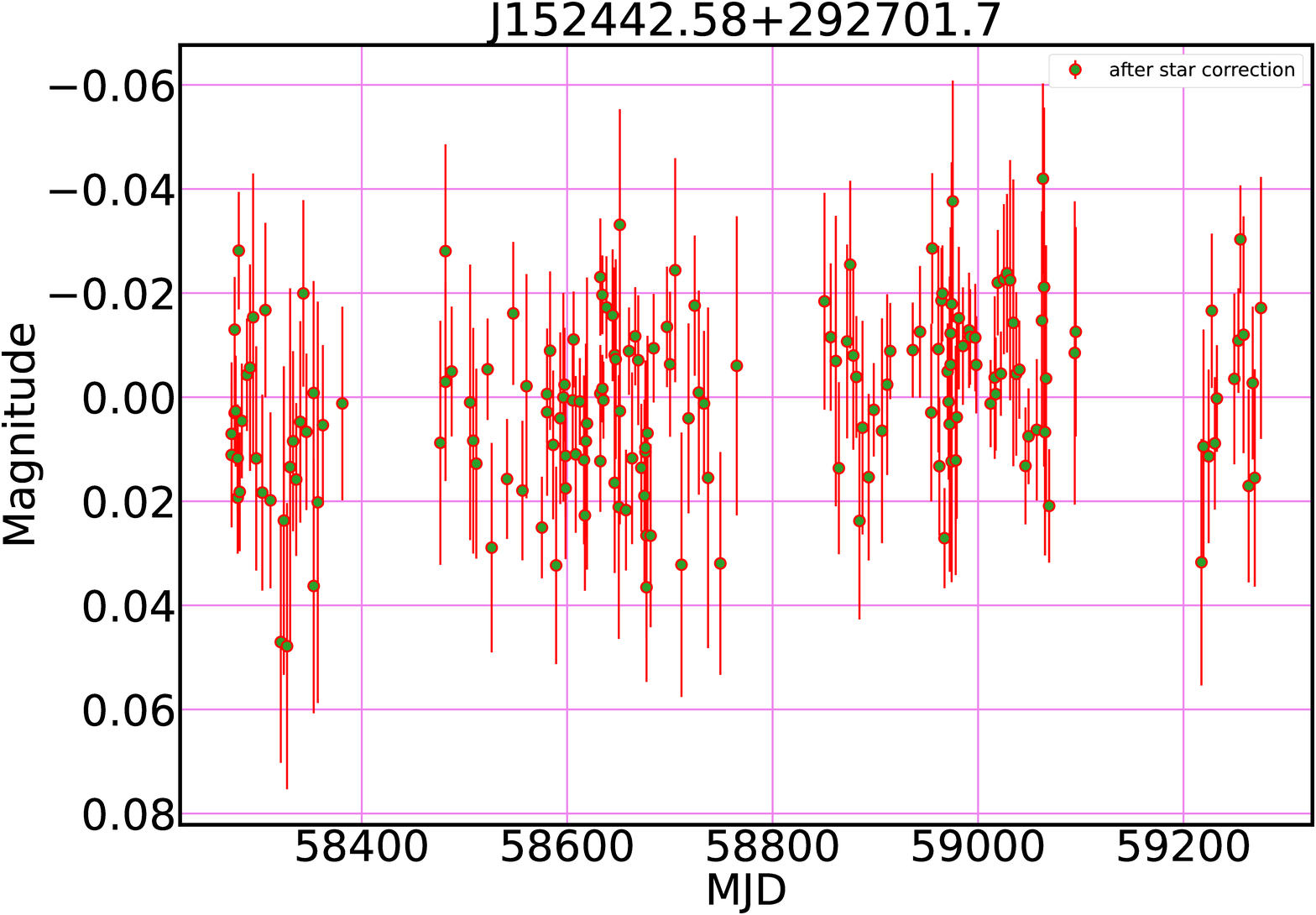}{a}{\emph{Top row:} J163159.59+243740.3 (left) and J152442.58+292701.7 (right) after filtering out the outliers shown by black star markers, color correction, zero point correction. \emph{Bottom row:} J163159.59+243740.3 (left) and J152442.58+292701.7 (right) after the reference star correction.}


For each ZTF photometric band we used the following criteria: (i) a photometric zero point (ZP) of a given image should not be higher than a ZP$_{\rm mean} - $0.2$\times$\emph{(airmass)} for the $g$ band (0.15$\times$\emph{(airmass)} for the $r$ band); (ii) a ZP RMS should be $<0.06$~mag; (iii) there should be at least 80 color calibrators in the $g$ band (or 120 in the $r$ band). 

\subsection{Color and zero-point corrections}

We perform a color ZP correction using the $g-r$ source color as described in the ZTF documentation. Then we apply a ZP correction by minimizing the $\chi^2$ statistics between data points originating from different 64 CCD quadrants in the ZTF detector.

We rank the quadrants by the number of points in them, and at the first step we calculate the $\chi^2$ statistics for the quadrant with the largest number of points. Next, we add measurements from the second most common quadrant and derive the ZP shift for this quadrant, which minimizes the combined $\chi^2$ statistics of the dataset. Then we repeat the algorithm iteratively for all quadrants, data points from which constitute the light curve. In other words, we derive such values of ZP shifts for individual quadrants so that the variability in a combined dataset is the lowest.
Such a technique can reduce a derived measure of real physical variability if a source was placed in different CCD quadrants at different variability phases (which is unlikely especially for stochastically variable sources), but this is perhaps the simplest approach to eliminate ZP deviations among CCD quadrants, which can introduce a spurious source variability. 

\subsection{Reference star correction}

The final step of light curves post-processing removes global trends originating from long-term ZP changes by using a ``median reference star''. We select up-to 50 bright nearby stars around the source of interest, which are marked as non-variable in the GAIA catalog and retrieve their light curves from the ZTF Forced Photometry Service. Each light curve is corrected using the filtering/correction as described above. 

Then we adjust the reference star light curves by applying the difference between the median magnitude of the source of interest (candidate AGN) and the median magnitude of each star and keep only those data points, which originate from the same ZTF images used for the construction of the AGN light curve. Next, we calculate the difference between the magnitude of the source of interest and the median value of the magnitude of all reference stars measured at the same image (after applying the median magnitude difference as described above). 

This simple correction allows us to correct flux measurements of an AGN candidate by the median reference star also removing potential variable stars, which are not flagged in GAIA.
Finally, we derive a fully corrected light curve of an AGN candidate, which can then be analyzed statistically.

The light curves before and after median reference star correction are shown in Fig~.\ref{a} (top and bottom rows). The object J163159.59+243740.2 (left panels) exhibits strong variability both on small timescales related to the size of the accretion disk and, accordingly, the mass of the black hole and on large timescales: here the median reference star correction does not change the variability assessment. However, for the object J152442.58+292701.7 the apparent variability seen after filtering spurious data points becomes less significant when we apply the median reference star correction. We can see that for a typical SDSS-selected AGN candidate ($g \sim 17-17.5$~mag) we can probe the variability at the 1--3\%\ level.











\section{Implications}



By filtering the outliers originating from data processing imperfections, correcting ZP differences and then correcting for systematic uncertainties using reference stars, one can reach the sensitivity of 1--3\%\ in the detection of stochastic variability of weak type-I AGN candidates using publicly available data from modern time-domain surveys such as ZTF. Currently, the bottleneck that limits the scalability of our analysis is that the ZTF Forced-Photometry Service can process no more than 100 objects at a time and it takes hours to weeks to generate light curves from difference images. In the future, when a full database with difference image light curves becomes available, we will be able to process tens of thousands of objects and complete the first census of weak AGN including those powered by IMBHs and ``light-weight'' SMBHs.

\acknowledgements MD acknowledges the ADASS 2021 organizing committee for providing financial aid. The authors are supported by the RSF grant 17-72-20119.

\bibliography{X4-003}  


\end{document}